\documentclass{llncs}

\usepackage{graphicx}
\usepackage{multicol,multirow,colortbl,booktabs}
\usepackage{graphicx}
\usepackage{cite}
\usepackage{comment}
\usepackage{xcolor}
\usepackage{tikz}
\usepackage{listings}
\usepackage{caption}
\usepackage{microtype}
\usepackage{subfigure}
\usepackage{ctable}
\usepackage{url}
\usepackage{wrapfig}
\usepackage[cmex10]{amsmath}
\usepackage[pdftex,
        colorlinks=false,
        pdftitle={Scalable Recursive State Compression},
        pdfauthor={Alfons Laarman},
        pdfsubject={},
        pdfkeywords={compression, model checking, tree-based algorithm},
        ]{hyperref}
        
\usepackage{amsfonts}
\title{Parallel Recursive State Compression for Free}

\author{Alfons Laarman, Jaco van de Pol, Michael Weber}

\institute{\email{\{a.w.laarman,vdpol,michaelw\}@cs.utwente.nl}\\
Formal Methods and Tools, University of Twente, The Netherlands
}

\newcommand\rmit[1]{\textit{\textrm{#1}}}

\lstdefinelanguage{simple}{
	morekeywords={for, proc, if, then, else, bool, int, true, false,
	return, type, parallel_for, while, to, inout},
	sensitive=false,
	morecomment=[l]{//},
	morecomment=[s]{/*}{*/}
}
\newcommand\state[1]{\langle{#1}\rangle}

\newcommand\floor[1]{\left \lfloor {#1} \right \rfloor}
\newcommand\ceil[1]{\left \lceil {#1} \right \rceil}

\newcommand\concept[1]{\textit{#1}}
\newcommand\COLLAPSE{\mbox{\scshape Collapse}}
\newcommand\SPIN{\mbox{\scshape Spin}}
\newcommand\PROMELA{\mbox{\scshape Promela}}
\newcommand\BEEM{\mbox{\scshape Beem}}

\newcommand{\oh}{\ensuremath{\mathcal{O}}}

\begin{document}

\maketitle

\begin{abstract}

This paper focuses on reducing memory usage in enumerative model checking, 
while maintaining the multi-core scalability obtained in earlier work.  We 
present a multi-core tree-based compression method, which works by 
leveraging sharing among sub-vectors of state vectors.

An algorithmic analysis of both worst-case and optimal compression ratios 
shows the potential to compress even large states to a small constant on 
average (8 bytes).  Our experiments demonstrate that this holds up in 
practice: the median compression ratio of 279 measured experiments is 
within 17\% of the optimum for tree compression, and five times better 
than the median compression ratio of \SPIN's \COLLAPSE\ compression.

Our algorithms are implemented in the LTSmin tool, and our experiments 
show that for model checking, multi-core tree compression pays its own 
way: it comes virtually without overhead compared to the fastest hash 
table-based methods.
\end{abstract}

\section{Introduction}
Many verification problems are computationally intensive tasks that 
can benefit from extra speedups. Considering recent hardware trends, 
these speedups do not come automatically for sequential exploration 
algorithms, but require exploitation of the parallelism within 
multi-core CPUs. In a previous paper, we have shown how to realize 
scalable multi-core reachability \cite{eemcs18437}, a basic task shared 
by many different approaches to verification. 

Reachability searches through all the \concept{states} of the program under
verification to find errors or deadlocks. 
It is bound by the number of states that fit into the main 
memory. Since states typically consist of large \concept{vectors} 
with one \concept{slot} for each program variable,
only small parts are updated for every step in the program.
Hence, storing a state in its entirety 
results in unnecessary and considerable overhead. State compression solves 
this problem, as this paper will show, at a negligible performance 
penalty and with better scalability than uncompressed hash tables.

\paragraph*{Related work.}

In the following, we identify compression techniques suitable for
(on-the-fly) enumerative model checking.
We distinguish between \textit{generic} and
\textit{informed} techniques.

\textit{Generic compression} methods, 
like Huffman encoding and run length encoding, 
have been considered for explicit state vectors with meager results 
\cite{670549, Geldenhuys}. These \concept{entropy encoding} methods
reduce \concept{information entropy} \cite{citeulike:308696}
by assuming common bit patterns. Such patterns have
to be defined statically and cannot be ``learned''
(as in dynamic Huffman encoding),
because the encoding may not change during state space 
exploration. Otherwise, desirable properties, like fast 
equivalence checks on states and constant-time state space inclusion 
checks, will be lost.

Other work focuses on efficient storage in hash tables
\cite{10.1109/TC.1984.1676499, 1767120}.
The assumption is that a uniformly distributed
subset of $n$ elements from the universe \textit U is stored in a
hash table. If each element in $U$ hashes to a unique location in the 
table, only one bit is needed to encode the presence of the element.
If, however, the hash function is not so perfect or $U$ is larger than
the table, then at least a quotient of the key needs to be stored and
collisions need to be dealt with.
This technique is therefore known as \concept{key quotienting}.
While its benefit is that the compression ratio is constant for any input
(not just constant on average), compression is only significant for
small universes\cite{1767120}, smaller than we encounter in model checking
(this universe consists of all possible combinations of the slot values,
not to be confused with the set of reachable states, which is typically much smaller).

The information theoretical lower bound on compression, or the 
\concept{information entropy},
can be reduced further if the format of
the input is known in advance (certain subsets of 
\textit U become more likely). 
This is what constitutes the class of \concept{informed compression} techniques. It includes works that
provide specialized storage schemes for certain specific state 
structures, like petri-nets \cite{Evangelista} or timed automata 
\cite{Larsen97efficientverification}. But, also \COLLAPSE\ compression 
introduced by Holzmann for the model checker \SPIN
\cite {Holzmann97statecompression}. It takes into account  
the independent parts of the state vector. Independent parts are 
identified as the global variables and the local variables belonging to 
different processes in the \SPIN-specific language \PROMELA.

Blom et al. \cite{so64523} present a more generic approach, 
based on a tree. All variables of a state 
are treated as independent and stored recursively in a binary tree of hash 
tables. The method was mainly used to decrease network traffic for 
distributed model checking. Like \COLLAPSE, this is a form of 
informed  compression, because it depends on the assumption that 
subsequent states only differ slightly.

\paragraph*{Problem statement.}
Information theory dictates that the more information we have on the data 
that is being compressed, the lower the entropy and the higher the 
achievable compression. Favorable results from informed 
compression techniques \cite{Evangelista, Larsen97efficientverification, 
Holzmann97statecompression, so64523} confirm this.
However, the techniques for petri-nets and timed automata employ specific 
properties of those systems (a deterministic transition relation and symbolic zone encoding respectively), and, therefore, 
are not applicable to enumerative model checking. 
\COLLAPSE\ requires local parts of 
the state vector to be syntactically identifiable and may thus not 
identify all equivalent parts among state vectors.
While tree compression showed more impressive compression ratios 
by analysis \cite{so64523} and is 
more generically applicable, it has never been benchmarked thoroughly 
and compared to other compression techniques nor has it been parallelized. 

Generic compression schemes can be added locally to a parallel 
reachability algorithm (see Sec.~\ref{s:background}).
They do not affect any concurrent 
parts of its implementation and even benefit scalability by lowering 
memory traffic \cite{670549}. 
While informed compression techniques can deliver better compression,
they require additional structures to record uniqueness of state vector 
parts.
With multiple processors constantly accessing these structures,
memory bandwidth is again increased and mutual exclusion locks are 
strained, thereby decreasing performance and scalability.
Thus the benefit of informed compression requires considerable design effort on modern multi-core CPUs with steep memory hierarchies.

Therefore, in this paper, we address two research questions: 
(1) does tree compression perform better than other 
state-of-the-art on-the-fly compression techniques
(most importantly \COLLAPSE),
(2) can parallel tree compression be implemented efficiently on multi-core 
CPUs.

\paragraph{Contribution.}
This paper explains a tree-based structure that 
enables high compression rates (higher than any other form of 
explicit-state compression that we could identify) 
and excellent performance. 
A parallel algorithm is presented (Sec.~\ref{sec:tree}) that makes this informed compression
technique scalable in spite of the multiple accesses to shared memory
that it requires, while also introducing \textit{maximal sharing}.
With an incremental algorithm, we further improve the performance,
reducing contention and memory footprint.

An analysis of compression ratios is provided (Sec.~\ref{sec:analysis}) 
and 
the results of extensive and realistic experiments
(Sec.~\ref{sec:experiments}) 
match closely to the analytical optima. 
The results also show that the incremental algorithm delivers excellent 
performance, even compared to uncompressed verification runs with a 
normal hash table. 
Benchmarks on multi-core machines show near-perfect scalability, even 
for cases which are sequentially already faster than the uncompressed run.


\section{Background}\label{s:background}

In Sec.~\ref{sec:reach}, we introduce
a parallel \concept{reachability} algorithm
using a shared hash table. 
The table's main functionality is the storage of a large set of state 
vectors of a fixed length~$k$. We call the elements of the vectors
\concept{slots} and assume that slots take  values from the integers,
possibly \concept{references} to complex values stored elsewhere
(hash tables or canonization techniques can be used to yield unique values for about any complex value).
Subsequently, in Sec.~\ref{sec:compression}, we explain two informed 
compression techniques that exploit similarity between different state 
vectors. While these techniques can be used to replace the hash table
in the reachability algorithm, they are are harder to parallelize as we
show in Sec.~\ref{sec:par}.

\subsection{Parallel Reachability}\label{sec:reach}
The parallel reachability algorithm (Alg.~\ref{alg:reach}) 
launches \textit{N} threads and assigns the initial states 
of the 
model under verification only to the \concept{open set} $S_{1}$
of the first thread (l.\ref{par1}). The open set can be 
implemented as a \concept{stack} or a 
\concept{queue}, depending on the desired 
search order (note that with $N>1$, the chosen search order will only be 
approximated, because the different threads will go through the search 
space independently).
The \concept{closed set} of visited states, 
\textit{DB}, is shared,
allowing threads executing the search algorithm
(l.\ref{search1}-\ref{search2})
to synchronize on the search space and each to 
explore a (disjoint) part of it\cite{eemcs18437}.
The \textsf{find\_or\_put} function returns \textit{true} when
\textit{succ} is found in \textit{DB}, and inserts it, when it is not.

Load balancing is needed so that workers that run out of work
(\hbox{$S_{id}=\emptyset$}) receive work from others. 
We implemented the function \textsf{load\_balance} 
as a form of Synchronous Random Polling \cite{sandersthesis}, 
which also ensures valid termination detection~\cite{eemcs18437}. 
It returns \textit{false} upon global termination.

\begin{center}
\vspace{-.9cm}
\begin{figure}[ht]
\begin{lstlisting}[caption={Parallel reachability algorithm with shared state storage}, label=alg:reach]
$S_1.$putall(initial_states) (*\label{par1}*)
parallel_for($\rmit{id}$ := 1 to N)
	while (load_balance($S_\rmit{id}$))
		work := 0
		while (work $<$ max $\land$ state := $S_\rmit{id}.$get()) (*\label{search1}*)
			count := 0
			for (succ $\in$ next_state(state))
				count := count + 1
				work := work + 1 (*\label{r:work}*)
				if ($\neg$find_or_put($\rmit{DB}$, succ)) then $S_\rmit{id}$.put(succ) 
			if (0 = count) then (*\textit{...report deadlock...} \label{search2}*)
\end{lstlisting}
\vspace{-1.8cm}
\end{figure}
\end{center}

\textit{DB} is generally implemented as a hash table. In
\cite{eemcs18437}, we presented a lockless hash table design, with which 
we were able to obtain almost perfect scalability.
However, with 16 cores, the physical memory, 64GB in our case, is filled 
in a matter of seconds, making memory the new bottleneck.
Informed compression techniques can solve this problem with an alternate
implementation of \textit{DB}.

\subsection{\COLLAPSE\ \& Tree Compression}\label{sec:compression}

\begin{figure}[htbp]
\vspace{-.6cm}
\begin{center}
\includegraphics[width=\linewidth]{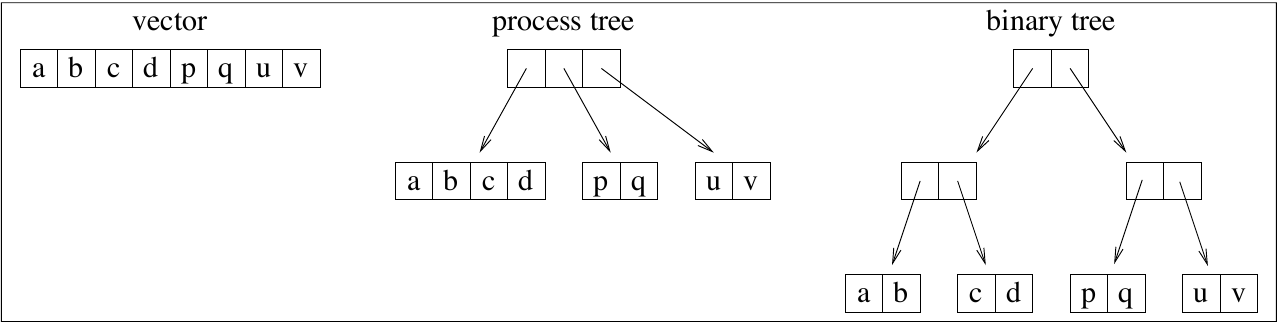}
\vspace{-.4cm}
\caption{Process table and (binary) tree for the system $X(a, b, c, d) 
\| Y(p, q) \| Z(u, v).$ Taken from \cite{Blom200368}.}
\label{fig:process_table}
\end{center}
\vspace{-.8cm}
\end{figure}

\COLLAPSE\ compression stores logical parts of the state vector 
in separate hash tables. 
A logical part is made up of state slots local to a 
specific process in the model, therefore the hash tables are called 
\concept{process tables}.
References to the parts in those process tables are then stored 
in a root hash table. Tree compression is similar, but works on the 
granularity of slots: tuples of slots are stored in hash tables at the
fringe of the tree,
which return a reference. References are then bundled as tuples
and recursively stored
in tables at the nodes of the binary tree.
Fig.~\ref{fig:process_table} shows the 
difference between the process tree and tree compression. 

\begin{figure}[]
\vspace{-.8cm}
\begin{center}
\includegraphics[width=\linewidth]{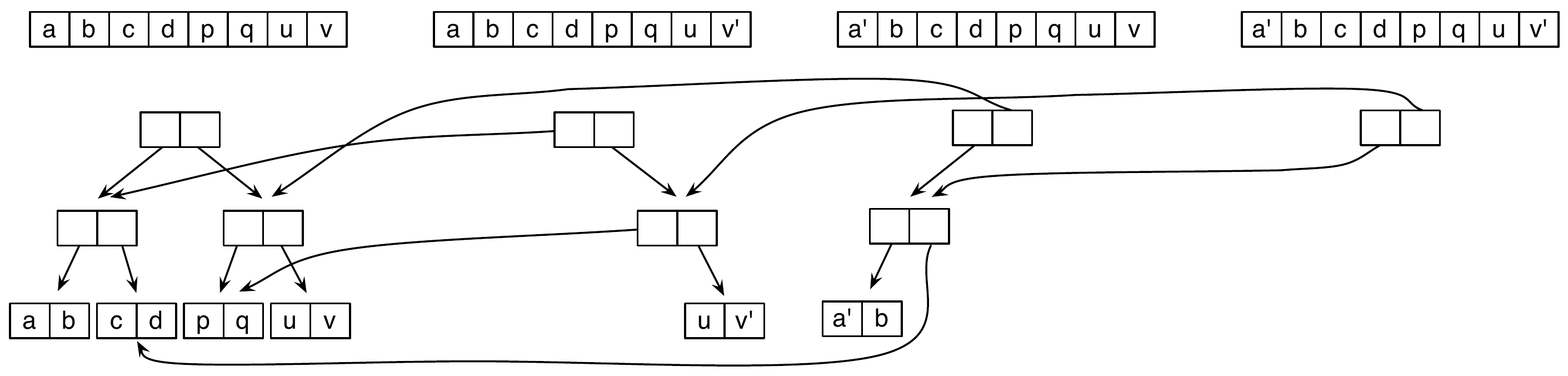}\label{fig:tree-basic}
\caption{Sharing of subtrees in tree compression}
\label{fig:sharing}
\end{center}
\vspace{-1.cm}
\end{figure}

When using a tree to store equal-length state vectors, compression is 
realized by the sharing of subtrees among entries. Fig. 
\ref{fig:sharing} illustrates this. Assuming that references have the 
same size as the slot values (say \textit b bits), we can determine the 
compression rate in this example.

Storing one vector in a tree, requires storing 
information for the extra tree nodes, resulting in a total of 
$8b + (4-1)\times 2b = 14b$ 
(not taking into account any implementation overhead from lookup 
structures). Each additional 
vector, however, can potentially share parts of the subtree with 
already-stored vectors.
The second and third, in the example, only require a total 
of $6b$ each and the fourth only $2b$. 
The four vectors would occupy $4 \times 8b = 32b$ when stored in 
a normal hash table. This gives a compression ratio of 
$28b / 32b = 7/8$, likely to improve with each additional vector that is 
stored. Databases that store longer vectors also achieve higher 
compression rates as we will investigate later.

\subsection{Why Parallelization is not Trivial}\label{sec:par}

Adding generic compression techniques to the above algorithm can be done 
locally by adding a line
\hbox{\textit{compr} := \textsf{compress}(\textit{succ})}
after l.\ref{r:work}, 
and storing \textit{compr} in \textit{DB}. 
This calculation in \textit{compress} only depends on \textit{succ} and is
therefore easy to parallelize.
If, however, a form of \emph{informed} compression is used, 
like \COLLAPSE\ or tree compression, the compressed value comes to depend 
on  previously inserted state parts, and the \textit{compress} 
function needs (multiple) accesses to the storage. 

\begin{wrapfigure}{r}{.6\linewidth}
  \vspace{-5ex}
  \includegraphics[width=\linewidth]{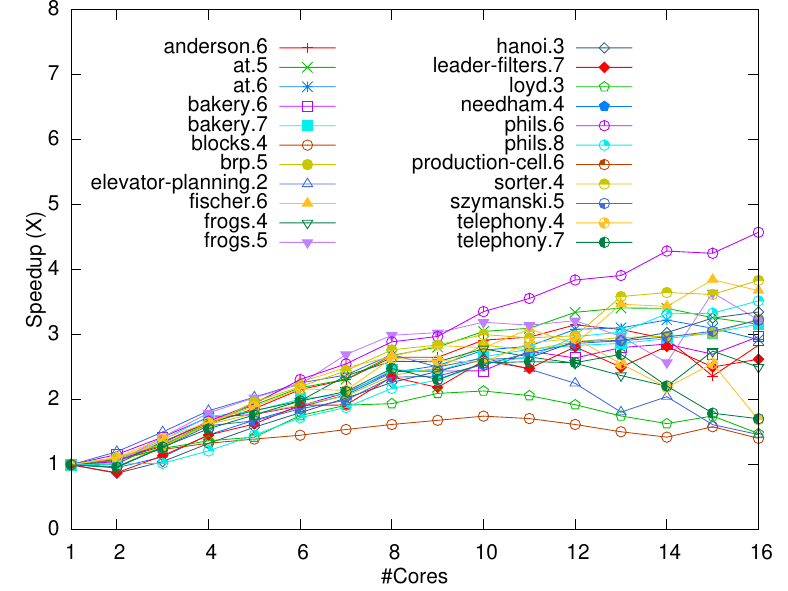}
  \vspace{-4.5ex}
    \caption{Speedup with \COLLAPSE.}\label{fig:speedupspin}
  \vspace{-6ex}
\end{wrapfigure}

Global locking or even locking at finer levels of granularity can be 
devastating for multi-core performance for single hash table lookups 
\cite{eemcs18437}. Informed compression algorithms, however, need 
multiple accesses and thus require careful attention when parallelized. 
Fig.~\ref{fig:speedupspin} shows that \SPIN's \COLLAPSE\ suffers from
scalability problems (experimental settings can be found in 
Sec.~\ref{sec:experiments}).

\section{Tree Database}\label{sec:tree}

Sec.~\ref{sec:alg} first describes the original tree compression algorithm 
from \cite{so64523}. In Sec.~\ref{sec:treepar}, 
\concept{maximal sharing} among tree nodes is introduced by merging the
multiple hash tables of the tree into a single fixed-size table. 
By simplifying the data structure in this way, we aid scalability.
Furthermore, we prove that it preserves consistency of the database's 
content. However, as we also show, the new tree will ``confuse'' tree 
nodes and erroneously report some vectors as \textit{seen}, while in fact they 
are \textit{new}. This is corrected by tagging root tree nodes, completing the parallelization.

Sec.~\ref{sec:refs} shows how tree references can also be used to compact
the size of the open set in Alg.~\ref{alg:reach}. Now that the necessary 
space reductions are obtained, the current section is concluded with
an algorithm that improves the performance of the tree database by using 
thread-local incremental information from the reachability search
(Sec.~\ref{sec:incr}).

\subsection{Basic Tree Database} \label{sec:alg}

The tuples shown in Fig.~\ref{fig:sharing} are stored in hash
tables, creating a \concept{balanced binary tree} of tables.
Such a tree has $k-1$ tree nodes, each of which has a number of siblings 
of both the left and the right subtree that is equal or off by one.
The \textsf{tree\_create} function in Alg.~\ref{alg:tree} generates the 
\textsf{Tree} structure accordingly, with \textsf{Node}s storing
\textit{left} and \textit{right} subtrees, a Table \textit{table} and the length of the (sub)tree $k$.

The \textsf{tree\_find\_or\_put} function takes as arguments a Tree and a 
state vector~$V$ (both of the same size $k>1$), and returns a tuple 
containing a 
reference to the inserted value and a boolean indicating whether the value 
was inserted before (\textit{seen}, or else: \textit{new}).
The function is recursively called on half of the state vector
(l.\ref{l:rec1}-\ref{l:rec2}) until the vector length is one.
The recursion 
ends here and a single value of the vector is returned.
At l.\ref{l:table}, the returned values of the left and right subtree 
are stored as a tuple in the hash table using the
\textsf{table\_find\_and\_put}
operation, which also returns a tuple containing a reference and a $seen/new$ 
boolean.

The function \textsf{lhalf} takes a vector $V$ as argument and returns the first
half of the vector: $\mathsf{lhalf}(V) = [V_0,\dots, V_{(\ceil{\frac k2}-1)}]$,
and symmetrically $\mathsf{rhalf}(V) = [V_{\ceil{\frac k2}},\dots, V_{(k-1)}]$. 
So, $|\mathsf{lhalf}(V)| = \ceil{|V|/2}$, and $| \mathsf{rhalf}(V)| = \floor{|V|/2}$. 

\newcommand{\uu}{\underline{\hspace*{1.5ex}}}

\begin{center}
\vspace{-.8cm}
\begin{figure}[ht]
\begin{lstlisting}[caption={Tree data structure and algorithm for the \textsf{tree\_find\_or\_put} function.}, label={alg:tree}]
type Tree = Node(Tree left, Tree right, Table table, int k) | Leaf(* \vspace{1ex} *)
proc Tree tree_create(k)
	if (k = 1)
		return Leaf
	return Node(tree_create($\ceil{\frac k2}$), tree_create($\floor{\frac k2}$), Table(2), k) (* \vspace{1ex} *)
proc (int, bool) tree_find_or_put(Leaf, $V$)
	return ($V[0]$, $\uu$) (* \vspace{1ex} *)
proc (int, bool) tree_find_or_put(Node(left, right, table, k), $V$)
	($R_{\textrm{left}}$, $\uu$) := tree_find_or_put(left, lhalf($V$)) (* \label{l:rec1} *)
	($R_{\textrm{right}}$, $\uu$) := tree_find_or_put(right, rhalf($V$)) (* \label{l:rec2} *)
	return table_find_or_put(table, $[R_{\textrm{left}}, R_{\textrm{right}}]$) (* \label{l:table} *)
\end{lstlisting}
\vspace{-1.7cm}
\end{figure}
\end{center}

\paragraph{Implementation requirements.}

A space-efficient implementation of the hash tables is crucial for
good compression ratios. Furthermore, resizing 
hash tables are required, because the unpredictable and widely varying 
tree node sizes (tables may store a crossproduct of their children 
as shown in Sec.~\ref{sec:analysis}). 
However, resizing replaces entries, in other words, it breaks
\concept{stable indexing}, thus making direct 
references between tree nodes impossible.
Therefore, in \cite{so64523}, stable indices were realized
by maintaining a second table with references.
Thus solving the problem, but increasing the number of cache misses and
the storage costs per entry by~50\%.

\subsection{Concurrent Tree Database}\label{sec:treepar}

Three conflicting requirements arise when attempting to parallelize 
Alg.~\ref{alg:tree}: 
(1) resizing is needed because the load of individual tables is 
unknown in advance and varies highly,
(2) stable indexing is needed, to allow for references to table entries, 
and
(3) calculating a globally unique index concurrently is costly, while
storing it requires extra memory as explained in the previous section.

An ideal solution would be to collapse all hash tables into a single 
non-resizable table. This would ensure stable indices without any 
overhead for administering them, while at the same time allowing
the use of a scalable hash table design \cite{eemcs18437}. Moreover, it 
will enable \concept{maximal sharing} of values between tree nodes,
possibly further reducing memory requirements.
\textit{But can all tree nodes safely be merged without corrupting the
contents of the database?}

To argue about consistency, we made a mathematical model of
Alg.~\ref{alg:tree} with one merged hash table.
The hash table uses stable indexing and is concurrent, hence each unique, 
inserted element will atomically yield one stable, unique index in the table.
Therefore, we can describe \textsf{table\_find\_or\_put} as a injective function:
$H_{k} : \mathbb{N}^{k} \rightarrow \mathbb{N}$. 
The \textsf{tree\_find\_or\_put} function can now be expressed as a 
recurrent relation ($T_{k} : \mathbb{N}^k \rightarrow \mathbb{N}$, for $k>1$ and $A\in \mathbb{N}^k$):\\
$T_{k}(A_0,\dots,A_{(k-1)}) = H_{2}(T_{\ceil{\frac k2}}(A_0,\dots,A_{(\ceil{\frac k2}-1)}),
			  T_{\floor{\frac k2}}(A_{\ceil{\frac k2}},\dots,A_{(k-1)}))$\\
$T_1(A_0) = A_0$.\\
We show that $T$ provides a similar injective function as $H$.

\paragraph{To prove (injection):} $C \equiv T_k(A) = T_k(B) \implies A=B$, with $A,B\in\mathbb{N}^k$.

Induction over $k$:

Base case: $T_1(x) = I(x)$, the identity function satisfies $C$ being 
injective. Assume $C$ holds $\forall i<k$ with $k>1$. We have to prove for all $A,B\in\mathbb{N}^k$, that:\\$
H_{2}(T_{\ceil{\frac k2}}(L(A)), T_{\floor{\frac k2}}(R(A)))=
H_2(T_{\ceil{\frac k2}}(L(B)), T_{\floor{\frac k2}}(R(B)))
\implies A=B$,\\ with $L(X) = X_0,\dots,X_{(\ceil{\frac k2}-1)}$ and
$R(X) = X_{\ceil{\frac k2}},\dots,X_{(k-1)}$. Note that:\\
$(*) \left \{\begin{array}{ll}
L(A) = L(B) \land R(A) = R(B) \}$ if $A=B\\
L(A) \neq L(B) \lor R(A) \neq R(B) \}$ if $A\neq B$. Hence,$\\
\end{array}
\right .
$\\
$\left .\begin{array}{l}

\left .\begin{array}{l}
T_k(A) = T_k(B) \implies
H_2(T_{\ceil{\frac k2}}(L(A)), T_{\floor{\frac k2}}(R(A)))=
H_2(T_{\ceil{\frac k2}}(L(B)), T_{\floor{\frac k2}}(R(B)))\\
\stackrel{inj. H_2}{\implies}
T_{\ceil{\frac k2}}(L(A)) = T_{\ceil{\frac k2}}(L(B)) \land
T_{\floor{\frac k2}}(R(A)) = T_{\floor{\frac k2}}(R(B))\\
\stackrel{ind. hyp.}{\implies}
L(A) = L(B) \land R(A) = R(B) \stackrel{(*)}{\implies} A=B\\
\end{array}\right .\\

\end{array}
\right .
$

Proving that $C$ holds for all $A$, $B$ and $k$. \qed

Now, it follows that an insert of a vector $A\in\mathbb{N}^k$ 
always yields a unique value for the root of the tree ($T_k$), thus
demonstrating that the contents of the tree database are 
not corrupted by merging the hash tables of the tree nodes.

However, the above also shows that Alg.~\ref{alg:tree} will not
always yield the right answer with merged hash tables.
Consider: $T_2(A_0, A_1) = H_2(0,0) = T_k(A_0,\dots,A_{(k-1)})$. In this case,
when the root node $T_k$ is inserted into $H$, it will return a boolean
indicating that the tuple $(0,0)$ was already seen, as it was inserted
for $T_2$ earlier.

\begin{center}
\vspace{-1.cm}
\begin{figure}[ht]
\begin{lstlisting}[caption={Data structure and algorithm for parallel \textsf{tree\_find\_or\_put} function.}, label={alg:treepar}]
type ConcurrentTree = CTree(Table table, int k) (*\vspace{.1cm}*)
proc (int, bool) tree_find_or_put(tree, $V$)
	$R$ := tree_rec(tree, $V$)
	$B$ := if CAS($R$.tag, non_root, is_also_root) then new else seen 	(*\label{c:cas}*)
	return ($R$, $B$) (*\vspace{1ex}*)
proc int tree_rec(CTree(table, k), $V$)
	if (k = 1)
		return $V[0]$
	$R_{\textrm{left}}$ := tree_rec(CTree(table, $\ceil{\frac k2}$), lhalf($V$))
	$R_{\textrm{right}}$ := tree_rec(CTree(table, $\floor{\frac k2}$), rhalf($V$))
	($R$, $\uu$) := table_find_or_put(table, $[R_{\textrm{left}}, R_{\textrm{right}}]$)
	return $R$
\end{lstlisting}
\vspace{-1.8cm}
\end{figure}
\end{center}

Nonetheless, we can use the fact that $T_k$ is an injection to create a 
concurrent tree database by adding one bit (a \textit{tag}) to the
merged hash table.
Alg.~\ref{alg:treepar} defines a new ConcurrentTree structure, only 
containing the merge \textit{table} and the length of the vectors $k$.
It separates the recursion in the \textsf{tree\_rec} function, which only 
returns a reference to the inserted node.
The \textsf{tree\_find\_or\_put} function now
atomically flips the tag on the entry (the tuple)
pointed to by $R$ in \textit{table}
from \textit{non\_root} to \textit{is\_also\_root}, if it was not
\textit{non\_root} before (see l.\ref{c:cas}).
To this end, it employs the hardware primitive
\concept{compare-and-swap} (\textsf{CAS}), which takes three arguments: a 
memory location (in this case, $R.\mathit{tag}$),
an \textit{old} value and a \textit{designated} 
value. CAS atomically compares the value \textit{val} at the memory location with 
\textit{old}, if equal, \textit{val} is replaced by \textit{designated}  
and true is returned, if not, false is returned.

\begin{wrapfigure}{r}{.2\linewidth}
\vspace{-1.3cm}
\begin{center}
\includegraphics[width=1\linewidth]{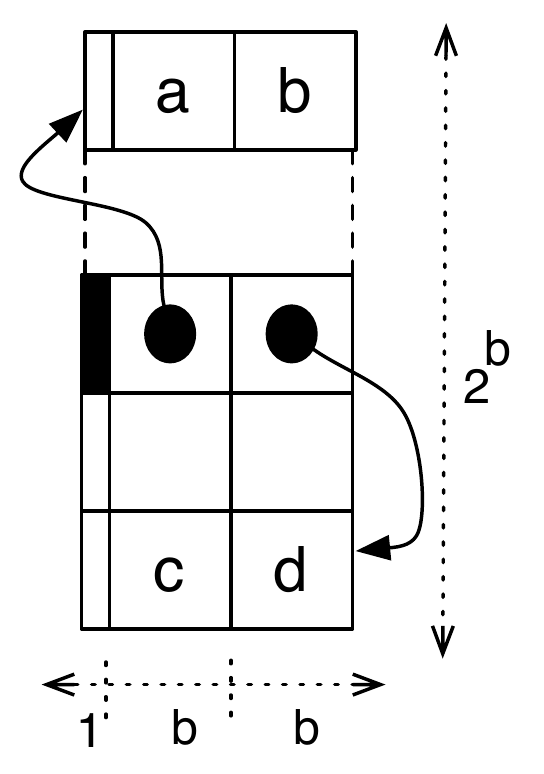}
\vspace{-.7cm}
\label{fig:tree-theoretical}
\caption{Memory layout for \textsf{CTree(Table, 4)} with $\state{a,b,c,d}$ 
inserted.}
\label{fig:mergedtable}
\end{center}
\vspace{-1.3cm}
\end{wrapfigure}

\paragraph{Implementation considerations.}
Crucial for efficient concurrency is \concept{memory layout}. While
a bit array or sparse bit vector may be used to implement the tags
(using $R$ as index), 
its parallelization is hardly efficient for high-throughput applications
like reachability analysis. Each modified bit will cause an entire
cache line (with typically thousands of other bits) to become
\concept{dirty}, causing other CPUs accessing the same memory 
region to be forced to update the line from main memory. 
The latter operation is multiple orders of 
magnitude more expensive than normal (cached) operations. Therefore, we 
merge the bit array/vector into the 
hash table \textit{table} as shown in Fig~\ref{fig:mergedtable},
for this increases the spatial locality of node
accesses with a factor proportional to the width of tree nodes.
The small column on the left represents the bit array with black
entries indicating \textit{is\_also\_root}. The appropriate size of $b$
is discussed in Sec.~\ref{sec:analysis}.

Furthermore, we used the lockless hash table 
presented in \cite{eemcs18437}, which normally uses \concept{memoized 
hashes} in order to speed up probing over larger keys.
Since the stored tree nodes
are relatively small, we dropped the memoize hashes,
demonstrating that this hash table design also functions well without
additional memory overhead.

\subsection{References in the Open Set}\label{sec:refs}

Now that tree compression reduces the space required for state 
storage, we observed that the open sets of the parallel reachability
algorithm can become a memory bottleneck \cite{ltsmin}. 
A solution is to store references 
to the root tree node in the open set as illustrated by
Alg.~\ref{alg:reachrefs}, which is a modification of
l.\ref{search1}-\ref{search2} from Alg.~\ref{alg:reach}.

\begin{center}
\vspace{-.8cm}
\begin{figure}[ht]
\begin{lstlisting}[caption={Reachability analysis algorithm with references in the open set.}, label=alg:reachrefs]
while (ref := $S_{id}.$get())
	state := tree_get(DB, ref)
	for (succ $\in$ next_state(state))
		(newref, seen) := tree_find_or_put(DB, succ)
		if ($\neg$seen) 
			$S_{id}$.put(newref)
\end{lstlisting}
\vspace{-1.6cm}
\end{figure}
\end{center}

The \textsf{tree\_get} function is shown in Alg.~\ref{alg:tree-get}. It 
reconstructs the vector from a reference. 
References are looked up in \textit{table} using the \textsf{table\_get} 
function, which returns the tuple stored in the table.
The algorithm recursively calls itself until $k=1$, 
at this point \textit{ref\_or\_val} is known to be a slot value and is 
returned as vector of size 1.
Results then propagate back up the tree and are concatenated on
l.\ref{l:concat}, until the full vector of length $k$ is restored at 
the root of the tree.

\begin{center}
\vspace{-.9cm}
\begin{figure}[ht]
\begin{lstlisting}[caption={Algorithm for tree vector retrieval from a reference}, label=alg:tree-get]
proc int$[]$ tree_get(CTree(table, k), val_or_ref)
	if (k = 1)
		return $[$val_or_ref$]$
	$[R_\textrm{left}$, $R_\textrm{right}]$ := table_get(table, val_or_ref)
	$V_\textrm{left}$ := tree_get(CTree(table, $\ceil{\frac k2}$), $R_\textrm{left}$)
	$V_\textrm{right}$ := tree_get(CTree(table, $\floor{\frac k2}$), $R_\textrm{right}$)
	return concat($V_\textrm{left}$, $V_\textrm{right}$) (* \label{l:concat} *)
\end{lstlisting}
\vspace{-1.7cm}
\end{figure}
\end{center}

\subsection{Incremental Tree Database}~\label{sec:incr}

The time complexity of the tree compression algorithm, measured in the 
number of hash table accesses, is linear in the
number of state slots. However, because of today's steep memory 
hierarchies these random memory accesses are expensive. Luckily, 
the same principle that tree compression exploits to deliver good 
state compression, can also be used to speedup the algorithm. The only
entries that need to be inserted into the node table are the slots that
actually changed with regard to the previous state and the tree paths 
that lead to these nodes. For a state vector of size $k$, the
number of table accesses can be brought down to $\log_{2}(k)$ (the height 
of the tree) assuming only one slot changed.
When $c$ slots change, the
maximum number of accesses is $c\times \log_2(k)$, 
but likely fewer if the slots are close to each other in the tree
(due to shared paths to the root).

Alg.~\ref{alg:incr} is the incremental variant of the 
\textsf{tree\_find\_or\_put} function.
The callee has to supply additional arguments: $P$ is the predecessor 
state of $V$ ($V\in \textsf{next\_state}(P)$ in Alg.~\ref{alg:reach}) and \textsf{RTree} is a 
\textsf{ReferenceTree} containing the balanced binary tree of references
created for $P$. 
\textsf{RTree} is also updated with the tree node references for~$V$. 
\textsf{tree\_find\_or\_put} needs to be adapted to pass 
the arguments accordingly.

\begin{center}
\vspace{-.9cm}
\begin{figure}[ht]
\begin{lstlisting}[caption={ReferenceTree structure and incremental \textsf{tree\_rec} function.}, label={alg:incr}]
type ReferenceTree = RTree(ReferenceTree left, ReferenceTree right, int ref) | Leaf (*\vspace{1ex}*)
proc (int, bool) tree_rec(CTree(table, k), $V$, $P$, Leaf)
	return ($V[0]$, $V[0]=P[0]$) (*\label{l:local}\vspace{1ex}*)
proc (int, bool) tree_rec(CTree(table, k), $V$, $P$, inout RTree(left, right, ref))
	($R_\textrm{left}$, $B_\textrm{left}$) := tree_rec(CTree(table, $\ceil{\frac k2}$), lhalf($V$), lhalf($P$), left)
	($R_\textrm{right}$, $B_\textrm{right}$) := tree_rec(CTree(table, $\floor{\frac k2}$), rhalf($V$), rhalf($P$), right)
	if ($\neg B_\textrm{left}\lor\neg B_\textrm{right}$) (*\label{l:cond1}*)
		(ref, $\uu$) := table_find_or_put$($table, $[R_\textrm{left}, R_\textrm{right}]$)  (*\label{l:cond2}*)
	return (ref, $B_\textrm{left}\land B_\textrm{right}$)
\end{lstlisting}
\vspace{-1.7cm}
\end{figure}
\end{center}

The boolean in the return tuple now indicates thread-local 
similarities between subvectors of $V$ and $P$ (see l.\ref{l:local}).
This boolean is used on l.\ref{l:cond1} 
as a condition for the hash table access;
if the left or the right subvectors are not the same, then \textsf{RTree} 
is updated with a new reference that is looked up in \textit{table}. 
For initial states, without predecessor states, the algorithm can be 
initialized with an imaginary predecessor state $P$ and tree
\textsf{RTree} containing reserved values, thus forcing updates.

\begin{wrapfigure}{r}{.55\linewidth}
\vspace{-1.cm}
\begin{center}
\includegraphics[width=\linewidth]{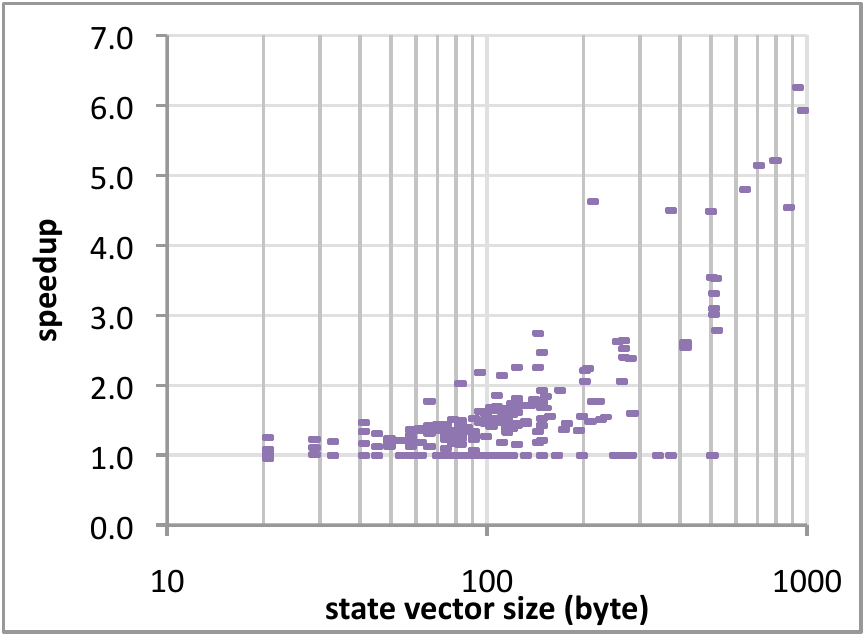}
\vspace{-.3cm}
\label{fig:tree-theoretical}
\caption{Speedup of Alg.~\ref{alg:incr} wrt. Alg.~\ref{alg:treepar}.}
\label{fig:noincr}
\end{center}
\vspace{-1.2cm}
\end{wrapfigure}

We measured the speedup of the incremental algorithm
compared to the original (for the experimental setup see
Sec.~\ref{sec:experiments}).
Fig.~\ref{fig:noincr} shows that the speedup is linearly dependent
on $\log(k)$, as expected. 

The incremental \textsf{tree\_find\_or\_put} function
changed its interface
with respect to Alg.~\ref{alg:treepar}. 
Alg.~\ref{alg:reachincr} presents a new search algorithm
(l.\ref{search1}-\ref{search2} in Alg.~\ref{alg:reach})
that also records the reference tree in the open set.
\textsf{RTree} \textit{refs} has become an input 
of the tree database, because it is also an output, it is copied to \textit{new\_refs}.

\begin{center}
\vspace{-.8cm}
\begin{figure}[ht]
\begin{lstlisting}[caption={Reachability analysis algorithm with incremental tree database.}, label=alg:reachincr]
while ((prev, refs) := $S_{id}.$get())
	for (next $\in$ next_state(prev))
		new_refs := copy(refs)
		($\uu$, seen) := tree_find_or_put($\textrm{DB}$, next, prev, new_refs)
		if ($\neg$seen) 
			$S_{id}$.put((next, new_refs))
\end{lstlisting}
\vspace{-.9cm}
\end{figure}
\end{center}

Because the internal tree node references are stored,
Alg.\ref{alg:reachincr}
increases the size of the open set by a factor of almost two.
To remedy this, either the \textsf{tree\_get} function
(Alg.~\ref{alg:tree-get}) can be adapted to also
return the reference trees, or the \textsf{tree\_get} function
can be integrated into the incremental algorithm
(Alg.~\ref{alg:incr}).
(We do not present such an algorithm due to space limitations.) 
We measured little slowdown due to the extra calculations and memory 
references introduced by the \textsf{tree\_get} algorithm
(about 10\% across a wide spectrum of input models). 

\section{Analysis of Compression Ratios}  \label{sec:analysis}

In the current section,
we establish the minimum and maximum compression ratio
for tree and \COLLAPSE\ compression. 
We count references and slots as stored in tuples at each tree node
(a single such \concept{node entry} thus has size 2).
We fix both references and slots 
to an equal size.\footnote
{For large tree databases references easily become 32 bits wide. This is 
usually an overestimation of the slot size.}

\paragraph{Tree compression.}
The worst case scenario occurs when storing a set of vectors $S$ 
with each $k$ identical slot values
($S = \{ \state{s,\dots,s}\mid s\in\{1,\dots,|S|\}\}$)
\cite{so64523}.
In this case, $n = |S|$ and storing each 
vector $v \in S$ takes $2(k - 1)$
($k-1$ node entries). 
The compression is: $(2(k-1)n) / (nk) = 2-2/k$. Occupying more tree 
entries is impossible, so always strictly less than twice the memory of 
the plain vectors is used.

Blom et al. \cite{so64523} also give an example that results in good tree 
compression: the storage of the cross product of a set of vectors 
$S = P \times P$, where P
consists of $m$ vectors of length $j=\frac12k$. 
The cross product ensures maximum reuse of the left and the right subtree, 
and results in $n = |S| = |P|^2= m^2$ entries in only the root node.
The left subtree 
stores $(j - 1 )|P|$ entries (taking naively the worst case), as does the 
right, resulting in a total of 
of $|S| + 2(j-1)|P|$ tree node entries. 
The size of the tree database for S becomes $2n + 2m(k-2)$.
The compression ratio is $2/k + 2/m - 4/(mk)$ (divide by $nk$), 
which can be approximated by $2/k$ for sufficiently large n (and hence m). 
Most vectors can thus be compressed to a size approaching
that of one node entry, which is logical since each new vector receives a 
unique root node entry (Sec.~\ref{sec:treepar}) and the other node entries 
are shared.

The optimal case occurs when all the individual tree nodes store 
cross products of their subtrees. This occurs when the 
value distribution is equal over all slots: 
$S = \{ \state{s_0,\dots, s_{k-1}}\mid s_i\in \{1,\dots,\sqrt[k]{n}\} \}$
and that $k=2^x$.
In this situation, the $\frac k2$ leaf nodes of the tree each receive 
$\sqrt[k/2]{n}$ entries: $\{ \state{s_i, s_{i+1}} \mid i = 2k \}$.
The nodes directly above the leafs, receive each the cross 
product of that as entries, etc, until the root node which receives 
$n$ entries (see Fig.~\ref{fig:tree-theoretical}). 

\begin{wrapfigure}{r}{.56\linewidth}
\vspace{-1.1cm}
\begin{center}
\includegraphics[width=1.02\linewidth]{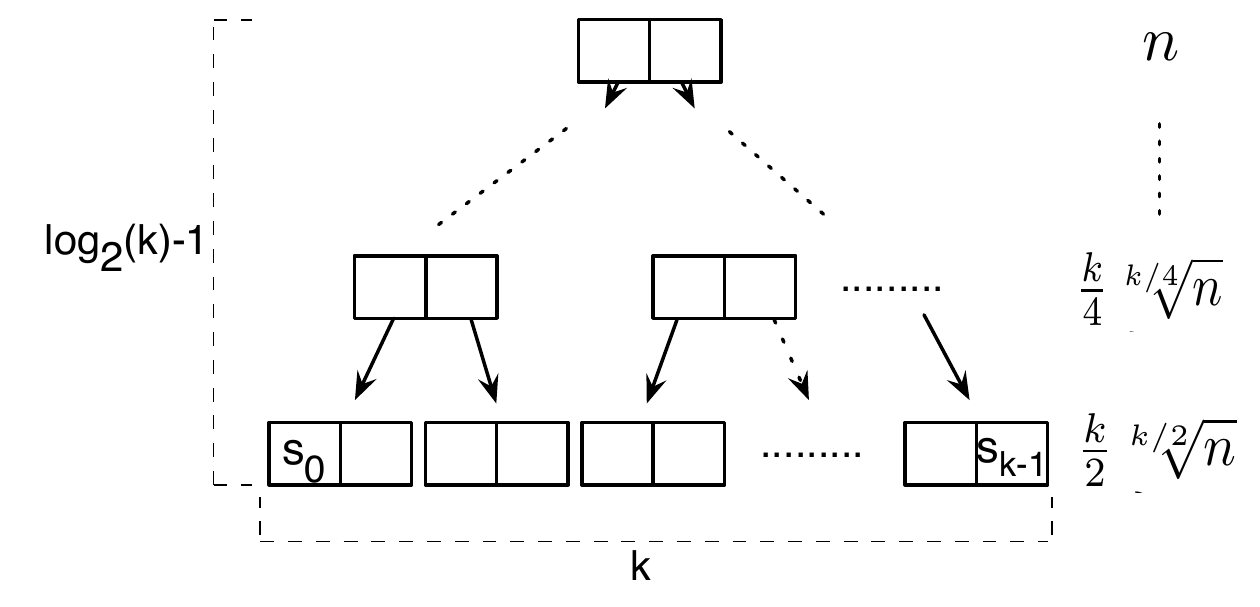}
\vspace{-.7cm}
\label{fig:tree-theoretical}
\caption{Optimal entries per tree node level.}
\label{fig:tree-theoretical}
\end{center}
\vspace{-1.cm}
\end{wrapfigure}

With this insight,
we could continue to calculate the total node entries for
the optimal case and try to deduce a smaller lower bound, but we can 
already see that the difference between the optimal case and the previous
case is negligible, since:
$n + \sqrt{n}(k-2) - (n + 2\sqrt{n} + 4\sqrt[4]{n} + 
\ldots (\log_2(k)$~times$)\ldots +
\frac 2k\sqrt[2/k]{n}) \ll n + \sqrt{n}(k-2)$,
for any reasonably large n and k.
From the comparison between the good and optimal case, we can conclude 
that only a cross product of entries in the root node is already 
near-optimal; 
the only way to get bad compression ratios may be when two related
variables are located at different halves of the state vector.

\paragraph{\COLLAPSE\ compression.}
Since the leafs of the process table are directly connected to the root, 
the compression ratios are easier to calculate.
To yield optimal compression for the process table, a more restrictive
scenario, than described for the tree above, needs to occur. 
We require $p$ symmetrical processes with each a local vector of $m$ slots 
($k = p \times m$). Related slots may only lay within the bounds of these
processes, take $S_m = \{\state{s,\dots,s}\mid s\in\{1,\dots,|S_m|\}\} $.
Each combination of different local vectors is inserted in the root table
(also if $S_m = \{ \state{s,1,\dots,1}\mid s\in\{1,\dots,|S_m|\}\}$),
 yielding $n = |S_{m}|^p$ root table entries. 
The total size of the process table becomes $pn+m\sqrt[p]{n}$. 
The compression ratio is
$ (pn + m\sqrt[p]{n}) / nk = \frac pk + m\frac{\sqrt[p]{n}}{nk}$. 
For large $n$ (hence $m$), the ratio approaches $\frac pk$.

\paragraph{Comparison.}
Tab. \ref{tab:theoretical} lists the achieved compression ratio for 
states, as  stored in a normal hash table, a process table and a tree 
database under the different scenarios that were sketched before. It 
shows that the worst case of the process table is not as bad as the worst 
case achieved by the tree. On the other hand, the best case scenario is
not as good as that from the tree, which compresses in this case
to a fixed constant. We also saw that the tree can reach near-optimal 
cases easily, placing few constraints on related slots (on the same half).
Therefore, we
can expect the tree to outperform the compression of process 
table in more cases, because the latter requires more restrictive 
conditions. Namely, related slots can only be within the fixed bounds
of the state vector (local to one process).

\begin{table}
\begin{center}
\vspace{-.8cm}
\caption{Theoretical compression ratios of \COLLAPSE\ and tree compression.}\label{tab:theoretical}
\begin{tabular}{|l|c|c|}
\hline
Structure & Worst case & Best case \\ \hline
Hash table \cite{eemcs18437}& $1$ 				& $1$ 		     \\[2px]
Process table				& $1+\frac pk$	& $\frac pk$ \\[2px]
Tree database (Alg.~\ref{alg:tree}, \ref{alg:treepar})	& $2 - \frac 2k$ 	& $\frac 2k$ \\[2px] \hline
\end{tabular}
	\vspace{-1.2cm}
\end{center}
\end{table}

\paragraph{In practice.}
With a few considerations, the analysis of this section
can be applied to both the parallel and the sequential tree databases:
(1) the parallel algorithm uses one extra \textit{tag} bit per node entry,
 causing insignificant overhead, and
(2) maximal sharing invalidates the worst-case analysis, but other sets
of vectors can be thought up to still cause the same worst-case size.
In practice, we can expect little gain from maximal sharing, since
the likelihood of similar subvectors decreases rapidly the larger these 
vectors are, while we saw that the most node entries are likely near the
top of the tree (representing larger subvectors).
(3) The original sequential version uses an extra reference per node entry 
of overhead (50\%!) to realize stable indexing (Sec.~\ref{sec:alg}). 
Therefore, the proposed concurrent tree implementation even improves the
compression ratio by a constant factor.


\section{Experiments}\label{sec:experiments}

We performed experiments on an AMD Opteron~8356 16-core 
($4 \times 4$ cores)
server with 64~GB RAM, running a patched Linux~2.6.32
kernel.\footnote{\url{https://bugzilla.kernel.org/show_bug.cgi?id=15618}, 
see also \cite{eemcs18437}}
All tools were compiled using {\sc gcc}~4.4.3 in 64-bit mode with high
compiler optimizations~(\texttt{-O3}).

We measured compression ratios and performance characteristics for the 
models of the \BEEM\ database \cite{beem} with three tools: DiVinE~2.2, 
\SPIN~5.2.5 
and our own model checker LTSmin \cite{eemcs18152,ltsmin}.
LTSmin implements Alg.~\ref{alg:treepar} 
using a specialized version of the hash table \cite{eemcs18437}
which inlines the $tags$ as discussed at the end of
Sec.~\ref{sec:treepar}.
Special care was taken to keep all parameters across the different model 
checkers the same. The size of the hash/node tables was fixed at $2^{28}$
elements to prevent resizing and model compilation options were optimized
on a per tool basis as described in earlier work~\cite{eemcs18152}.
We verified state  and transition counts with the \BEEM\ database and
DiVinE 2.2. The complete results with over 1500 benchmarks are available 
online\cite{spreadsheet}.

\subsection{Compression Ratios}

For a fair comparison of compression ratios between \SPIN\ and LTSmin, 
we must take into account the differences between the tools. 
The \BEEM\ models 
have been written in DVE format (DiVinE) and translated to \PROMELA.
The translated \BEEM\ models that \SPIN\ uses may have a different state 
vector length. 
LTSmin reads DVE inputs directly, but uses a standardized internal state 
representation with one 32-bit integer per \concept{state slot} (state variable) 
even if a state variable could be represented by a single byte.
Such an approach was chosen in order to reuse the model checking 
algorithms for other model inputs (like mCRL, mCRL2 and DiVinE 
\cite{bridging}). 
Thus, LTSmin can load \BEEM\ models  
directly, but blows up the state vector by an average factor of three. 
Therefore, we compare the average compressed state vector size instead of 
compression ratios.
\newcommand\tc[1]{\multicolumn{1}{c}{#1}}
\newcommand\cb[0]{\rowcolor{black!5}}
  
\begin{table}[htdp]
\vspace{-.9cm}
\begin{minipage}[b]{\linewidth}
  \caption{Original and compressed state sizes and memory usage for LTSmin 
    with hash table (\textit{Table}), \COLLAPSE\ (\textit{\SPIN}) and
    our tree compression (\textit{Tree}) for a representative selection of all 
    benchmarks.}

\vspace{-.2cm}
\label{tab:compression}
\begin{center}
  \tabcolsep=1em\renewcommand\arraystretch{1.05}
{\scriptsize\begin{tabular}{lrrrrrrr}
\toprule
\tc{\multirow{2}{*}{Model}} & 
\multicolumn{2}{c}{Orig. State [Byte]} &
\multicolumn{2}{c}{Compr. State [Byte]} &
\multicolumn{3}{c}{Memory [MB]} \\

\cmidrule(r){2-3}  \cmidrule(r){4-5}	\cmidrule(r){6-8}
& \SPIN & Tree
& \SPIN & Tree 
& Table\footnote{The hash table size is calculated on the base of the LTSmin state sizes} & \SPIN & Tree \\

\midrule\cb

\verb"at.6"	   & 68  & 56  & 36.9  & 8.0  & 8,576 & 4,756 & 1,227 \\
\verb"iprotocol.6" & 164 & 148 & 39.8  & 8.1  & 5,842 & 2,511 & 322  \\\cb
\verb"at.5"	   & 68  & 56  & 37.1  & 8.0  & 1,709 & 1,136 & 245  \\
\verb"bakery.7"	   & 48  & 80  & 27.4  & 8.8  & 2,216 & 721  & 245  \\\cb 
\verb"hanoi.3"	   & 116 & 228 & 112.1 & 13.8 & 3,120 & 1,533 & 188  \\
\verb"telephony.7" & 64  & 96  & 31.1  & 8.1  & 2,011 & 652  & 170  \\\cb
\verb"anderson.6"  & 68  & 76  & 31.7  & 8.1  & 1,329 & 552  & 140  \\
\verb"frogs.4"	   & 68  & 120 & 73.2  & 8.2  & 1,996 & 1,219 & 136  \\\cb
\verb"phils.6"	   & 140 & 120 & 58.5  & 9.3  & 1,642 & 780  & 127  \\
\verb"sorter.4"	   & 88  & 104 & 39.7  & 8.3  & 1,308 & 501  & 105  \\\cb
\verb"elev_plan.2"  & 52  & 140 & 67.1  & 9.2  & 1,526 & 732  & 100  \\
\verb"telephony.4" & 54  & 80  & 28.7  & 8.1  &   938 & 350  & 95   \\ \cb
\verb"fischer.6"   & 92  & 72  & 43.7  & 8.4  &   571 & 348  & 66   \\

\bottomrule

\end{tabular}}\vspace{-.23cm}
\end{center}
\end{minipage}\vspace{-.6cm}
\end{table}

Table~\ref{tab:compression} shows the uncompressed and compressed
vector sizes for \COLLAPSE\ and tree compression.  Tree
compression achieves better and almost constant state compression than \COLLAPSE\ for these selected models, even though
original state vectors are larger in most cases. This confirms the
results of our analysis.

We also measured peak memory usage for full state space exploration.
The benefits with respect to hash tables can be staggering for both
\COLLAPSE\ and tree compression: while the hash table column is in the
order of gigabytes, the compressed sizes are in the order of hundreds
of megabytes. An extreme case is \texttt{hanoi.3}, where tree
compression, although not optimal, is still an order of magnitude
better than \COLLAPSE\ using only 188~MB compared to 1.5~GB with \COLLAPSE\ 
and 3~GB with the hash table.

\begin{figure}
\vspace{-.6cm}
\begin{center}
\includegraphics[width=\linewidth]{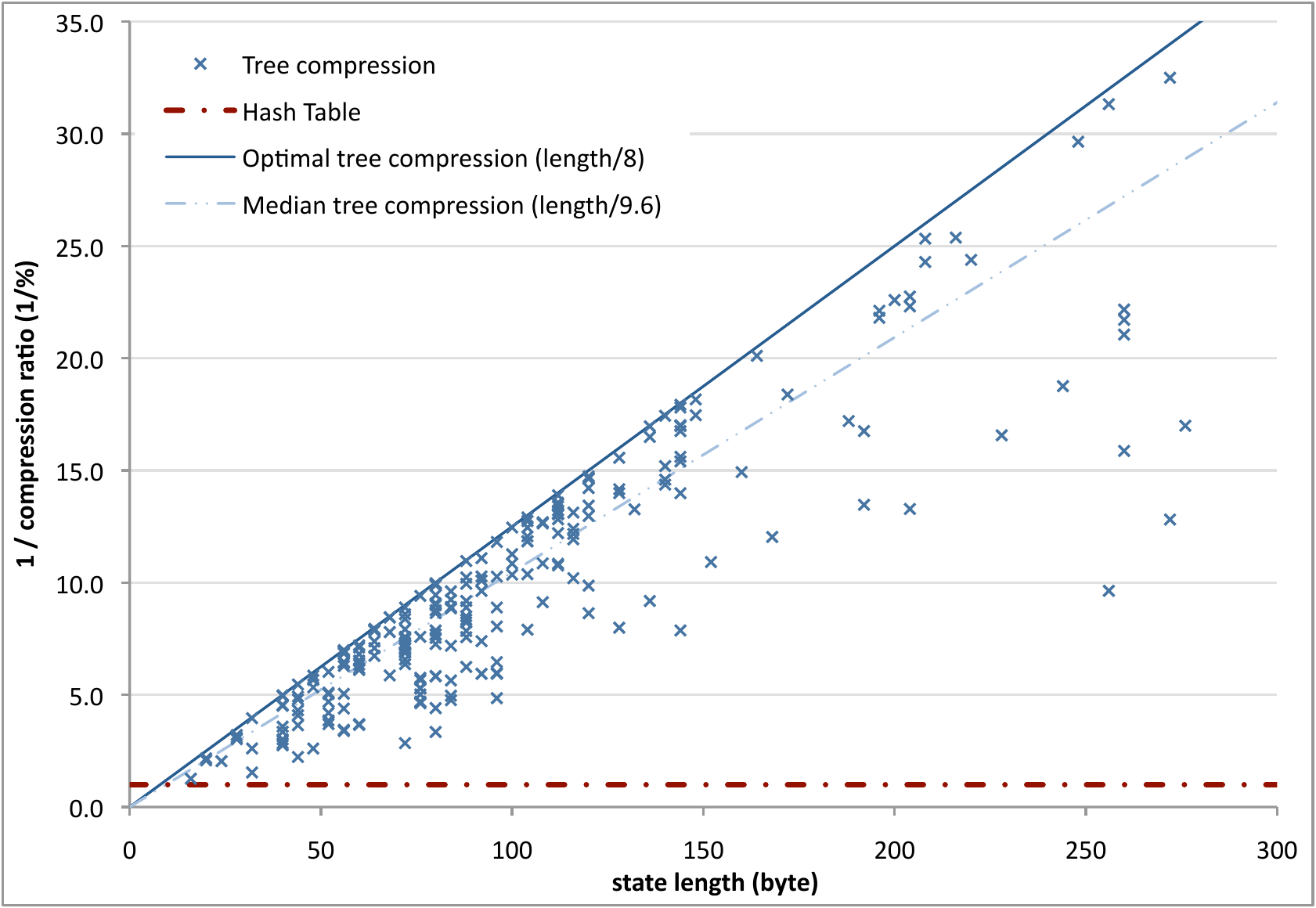}
\caption{Compression ratios for 279 models of the \BEEM\ database
  are close to optimal for tree compression.}
\label{fig:tree_all}
\end{center}
\vspace{-1.cm}
\end{figure}

To analyze the influence of the model on the compression ratio, we
plotted the inverse of the compression ratio against the state length
in Fig.~\ref{fig:tree_all}.
The line representing optimal compression
is derived from the analysis in Sec.~\ref{sec:analysis}
and is linearly dependent on the state size  (the average
compressed state size is close to 8~bytes: two 32-bit integers for
the dominating root node entries in the tree).

With tree compression, a total of 279~\BEEM\ models could each be fully
explored using a tree database of pre-configured size, never
occupying more than 4~GB memory.  Most models exhibit compression
ratios close to optimal; the line representing the median compression
ratio is merely 17\% below the optimal line.  The worst cases, with a
ratio of three times the optimal, are likely the result of
combinatorial growth concentrated around the center of the tree,
resulting in equally sized root, left and right sibling tree nodes.
Nevertheless, most sub-optimal cases lie close too half of the
optimal, suggesting only one ``full'' sibling of the root
node. (We verified this to be true for several models.)

\begin{figure}[tb]
  \parbox[t]{0.48\linewidth}{\centering
  	\includegraphics[width=\linewidth]{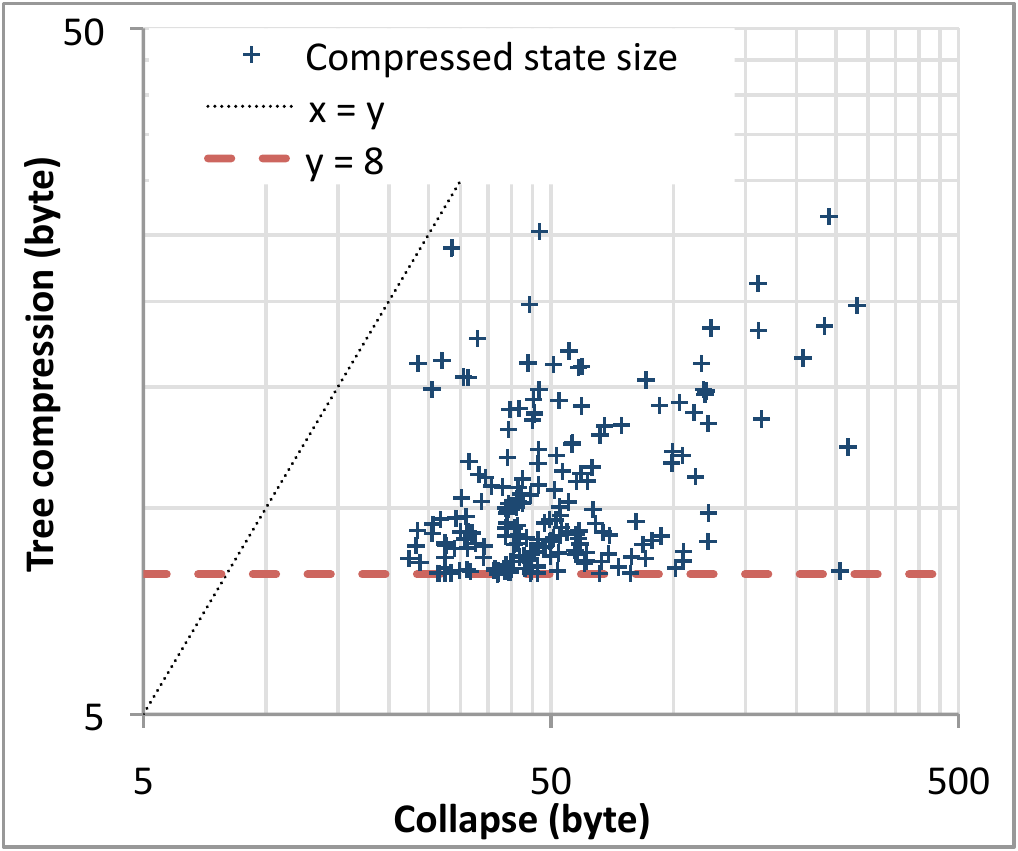}
    \caption{Log-log scatter plot of \COLLAPSE\ and tree-compressed state sizes
      (smaller is better): for all tested models, tree compression
      uses less memory.\label{fig:comp_all}}
  }
  \hspace{0.01\linewidth}
  \parbox[t]{0.48\linewidth}{\centering
    \includegraphics[width=\linewidth]{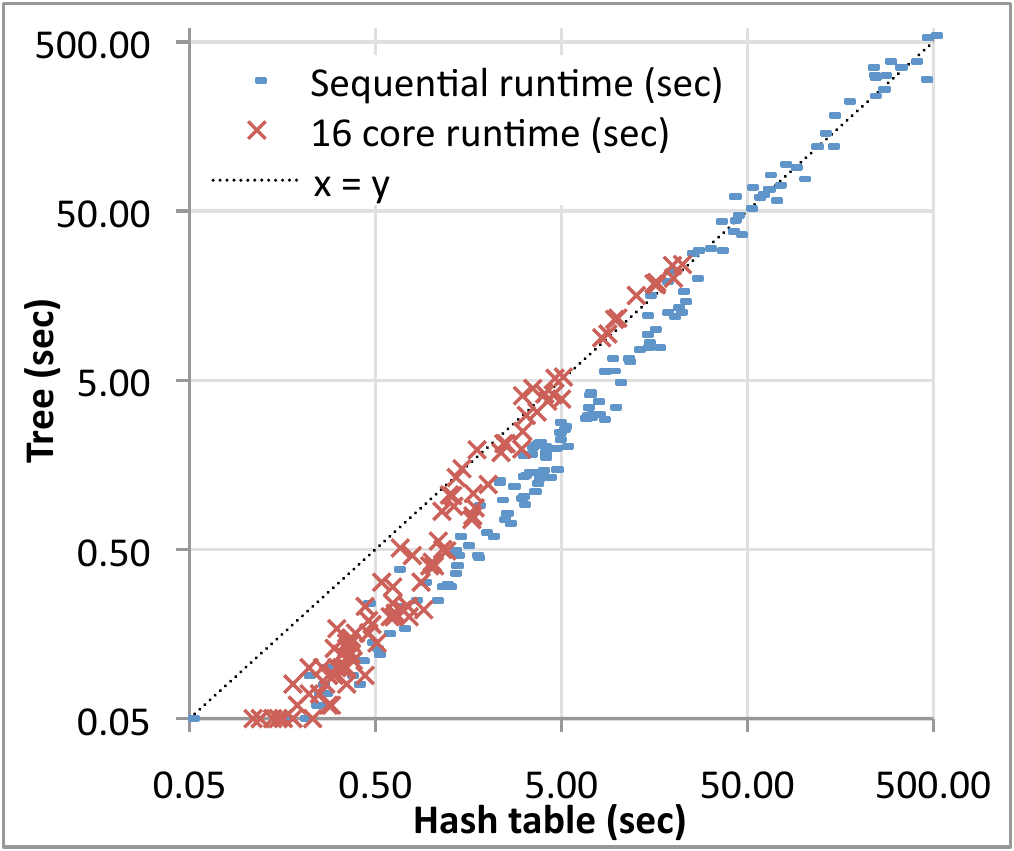}
    \caption{Log-log scatter plot of LTSmin run-times for state space
      exploration with either a hash table or tree
      compression.\label{fig:comp_all_runtime}}
  }
  \vspace{-3ex}
\end{figure}

Fig.~\ref{fig:comp_all} compares compressed state size of \COLLAPSE\ and
tree compression.  (We could not easily compare compressed state
\emph{space} sizes due to differing number of states for some models).
Tree compression performs better for all models in our data
set.  In many cases, the difference is an order of magnitude.  While
tree compression has an optimal compression ratio that is four times
better than \COLLAPSE's (empirically established), the median is even
five times better for the models of the \BEEM\ database. Finally,
as expected (see Sec.~\ref{sec:analysis}), we measured insignificant gains 
from the introduced maximal sharing. 

\subsection{Performance \& Scalability}

\begin{figure}[tp]
\vspace{-.4cm}
\hspace{-.2cm}
\subfigure[Total runtime] {\includegraphics[width=.52\linewidth]{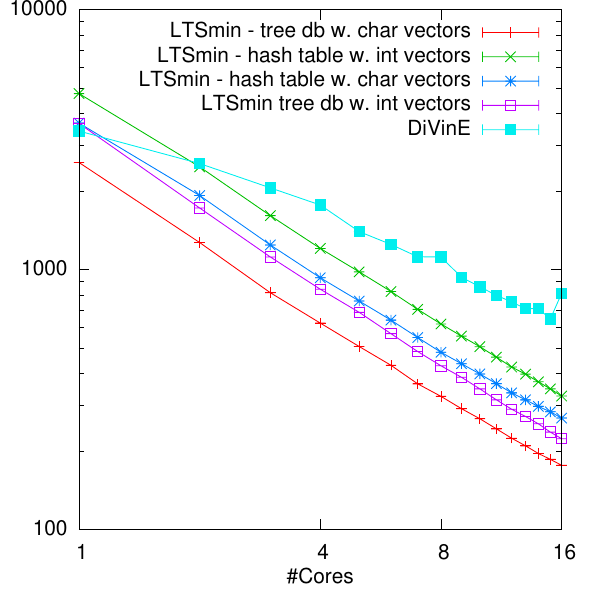}
\label{fig:runtime}}
\hspace{-.4cm}
\subfigure[Average (absolute) speedup] {\includegraphics[width=.52\linewidth]{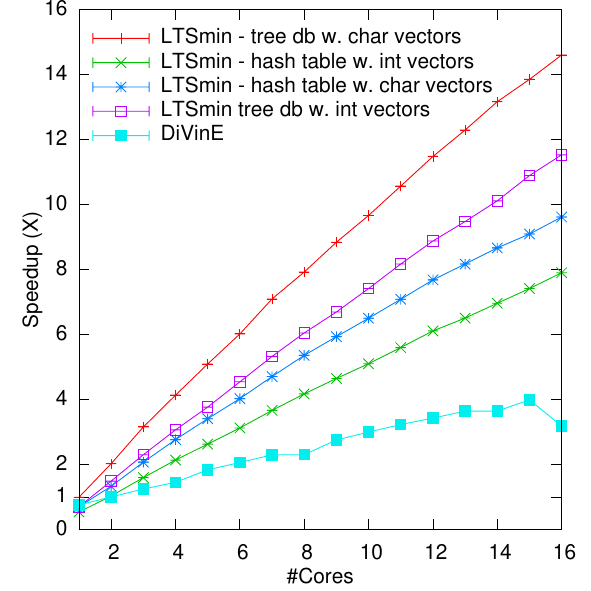}
\label{fig:speedup}}
\vspace{-.3cm}
\caption{Performance benchmarks for 158 models  
with DiVinE (hash table) and with LTSmin using tree compression and hash table.}
\label{fig:performance}
\vspace{-.6cm}
\end{figure}

We compared the performance of the tree database with a hash table in
DiVinE and LTSmin.  A comparison with \SPIN\ was already provided in
earlier work~\cite{eemcs18437}.  For a fair comparison, we modified a
version of LTSmin\footnote{this experimental version is distributed
  separately from LTSmin, because it breaks the language-independent
  interface.} to use the (three times) shorter state vectors
(\concept{char vectors}) of DiVinE directly. Fig.~\ref{fig:performance}
shows the total runtime of 158~\BEEM\ models, which fitted in machine
memory using both DiVinE and LTSmin. On average the run-time performance of tree compression is close to a hash table-based search
(see Fig.~\ref{fig:runtime}). 
However, the absolute speedup in Fig.~\ref{fig:speedup} shows that 
scalability is better with tree compression,
due to a lower memory footprint.

Fig.~\ref{fig:comp_all_runtime} compares the sequential and multi-core
performance of the fastest hash table implementation (LTSmin lockless
hash table with char vectors) with the tree database (also with char
vectors). The tree matches the performance of the hash table closely.

For both, sequential and multi-core, the performance of the tree
database is nearly the same as the fastest hash table implementation,
however, with significantly lower memory utilization.  For models with
fewer states, tree database performance is better than a hash table,
undoubtedly due to better cache utilization and lower memory bandwidth.


\section{Conclusions}

First, this paper presented an analysis and experimental evaluation of the 
compression ratios of tree compression and \COLLAPSE\ compression,
both informed compression techniques that are applicable in on-the-fly
model checking. Both analysis and experiments can be considered 
an implementation-independent comparison of the two techniques.
\COLLAPSE\ compression was considered the
state-of-the-art compression technique for enumerative model checking.
Tree compression was not evaluated as such before.
The latter is shown here to
perform better than the former, both analytically and in practice.
In particular, the median compression ratio of tree
compression is five times better than that of \COLLAPSE\ on the \BEEM\ 
benchmark set.  We consider this result representative to real-world
usage, due to the varied nature of the \BEEM\ models: the set
includes models drawn from extensive case studies on protocols and control 
systems, and, implementations of planning, scheduling and mutual exclusion algorithms
\cite{beemwebsite}.

Furthermore, we presented a solution for parallel tree compression
by merging all tree-node
tables into a single large table, thereby realizing maximal sharing 
between entries in these tables. This single hash table design even saves 
50\% in memory because it exhibits the required stable indexing without 
any bookkeeping.
We proved that the consistency is maintained and use only one bit per 
entry to parallelize tree insertions. 
Lastly, we presented an incremental tree compression algorithm
that requires a fraction of the table accesses
(typically $\oh(\log_{2}(k))$, i.e., logarithmic in the length of a state
vector), compared to the original algorithm.

Our experiments show that the incremental and parallel tree database
has the same performance as the hash table solutions in both LTSmin
and DiVinE (and by implication \SPIN~\cite{eemcs18437}).  Scalability
is also better.  All in all, the tree database provides a win-win
situation for parallel reachability problems. 

\paragraph*{Discussion.}
The absence of resizing could be considered a limitation in certain 
applications of the tree database. In model checking, however, we may 
safely dedicate the vast majority of available memory of a system to
the state storage.

The current implementation of LTSmin~\cite{website} supports a maximum
of $2^{32}$ tree nodes, yielding about $4\times10^{9}$ states with
optimal compression. In the future, we aim to
create a more flexible solution that can store more states and
automatically scales the number of bits needed per entry, depending on
the state vector size.
What has hold us back thus far from
implementing this are low-level issues, i.e., the ordering of
multiple atomic memory accesses across cache line boundaries behave
erratically on certain processors.

While this paper discusses tree compression mainly in the context of
reachability, it is not limited to this context. For example, on-the-fly 
algorithms for the verification of liveness properties can also benefit
from a space-efficient storage of states as demonstrated by \SPIN\ 
with its \COLLAPSE\ compression.

\paragraph*{Future Work.}
A few options are still open to improve tree compression.  The small
tree node entries cover a limited universe of values: $1 +
2\times \log_2(n)$.  This is an ideal case to employ
\concept{key quotienting} using
\concept{Cleary}~\cite{10.1109/TC.1984.1676499} or
\concept{Very Tight Hashtables}~\cite{1767120}.
Neither of the two techniques has been parallelized as far as we can tell.

Static analysis of the dependencies between transitions and state
slots could be used to reorder state slots and obtain a better
balanced tree, and hence better compression (see Sec.~\ref{sec:analysis}). 
Much like the variable ordering problem of BDDs \cite{136043}, finding the 
optimal reordering is an
exponential problem (a search through all permutations). While, we are
able to improve most of the worse cases by automatic
variable reordering, we did not yet find a good heuristic
for at least all \BEEM\ models.

Finally, it would be interesting to generalize the tree database by
accommodating for the storage of vectors of different sizes.


\subsection*{Acknowledgements}
We thank Elwin Pater for taking the time to proofread this work and 
provide feedback. We thank Stefan Blom for the many useful ideas 
that he provided.

\bibliography{main}
\bibliographystyle{plain}

\end{document}